\documentclass[12pt]{article}
\pdfoutput=1
\usepackage{latexsym}\usepackage{epsfig,amssymb,euscript}
\usepackage{amsmath} \usepackage{bbm,slashed}
\usepackage{color}
\usepackage{ulem}
\usepackage{cite}
\usepackage{graphicx}
\usepackage{hyperref}

\newcommand{\nn} {\nonumber\\}

\def\tilde{\widetilde}
\def\CO{\mathcal O}
\def\CL{\mathcal L}

\catcode`@=12

\numberwithin{equation}{section}

\begin{document}
\begin{titlepage}

\def\thefootnote{\fnsymbol{footnote}}

\begin{center}
\vskip -10pt
{\LARGE
{\bf
A Note on Holographic Non-Relativistic\\
\vskip 5pt
 Goldstone Bosons 
}
}
\end{center}

\bigskip
\begin{center}
{\large 
Riccardo Argurio,$^1$ Andrea Marzolla,$^1$\\ 
\vskip 5pt  Andrea Mezzalira$^{2,3}$
and Daniel Naegels$^1$}
\end{center}

\renewcommand{\thefootnote}{\arabic{footnote}}

\begin{center}
\vspace{0.2cm}
$^1$ {Physique Th\'eorique et Math\'ematique and International Solvay Institutes \\ Universit\'e Libre de Bruxelles, C.P. 231, 1050 Brussels, Belgium\\}
$^2$ {Institute Lorentz for Theoretical Physics, Leiden University\\P.O. Box 9506, Leiden 2300RA, The Netherlands\\}
$^3$ {School of Mathematics, Trinity College, Dublin 2, Ireland\\}
\vskip 5pt

\end{center}

\noindent
\begin{center} {\bf Abstract} \end{center}
\noindent
We consider a holographic set-up where relativistic invariance is broken by a chemical potential, and a non-abelian internal symmetry is broken spontaneously.  We use the tool of holographic renormalization in order to infer what can be learned purely by analytic boundary considerations. We find that  the expected Ward identities are
correctly reproduced. In particular, we obtain the identity which implies the non-commutation of a pair of broken charges, which leads to 
the presence of
Goldstone bosons with quadratic dispersion relations.

\end{titlepage}

\setcounter{footnote}{0}

\tableofcontents

\section{Introduction}

Goldstone's theorem has been originally proved in a Lorentz-invariant formulation. Its extension to non-relativistic settings is not straightforward and raises a problem of counting, as it was quite soon realized \cite{Nielsen:1975hm}: the number of massless modes does not match the number of broken generators in this context, and moreover the dispersion relation of non-relativistic Goldstone bosons is not any more constrained to be linear and with constant of proportionality fixed to the speed of light, as in the Lorentz-invariant case. Non-relativistic set-ups occur quite often in condensed matter physics, but the issue of the generalization of Goldstone's theorem to these set-ups has obviously great theoretical interest and only recently some main developments have been achieved in its understanding\cite{Watanabe:2011ec,Watanabe:2012hr,Kapustin:2012cr,Watanabe:2014fva}. The occurrence of Goldstone bosons with quadratic dispersion relation has been related to the presence of pairs of broken generators whose commutator has a non-trivial vacuum expectation value (VEV), which is forbidden in the case of Lorentz invariance. Such Goldstone bosons with quadratic dispersion relations are usually accompanied by a massive partner whose mass is proportional to the amount of Lorentz breaking. Eventually, there can still be massless particles which have linear dispersion relations, as in the relativistic case, but with a model dependent velocity. 

The aim of this note is to explore in a holographic set-up \cite{Maldacena:1997re,Witten:1998qj,Gubser:1998bc} what can be learned on these different kinds of light modes. In particular, we will focus on what can be extracted {\it a priori} in a given model, i.e. just by specifying how the symmetries are broken. The holographic model is supposed to represent a theory with a large number of degrees of freedom and at strong coupling, which displays a pattern of symmetry breaking allowing for the various types of Goldstone bosons to be present. We will study a model with the minimal requirements in order to expect all these light modes, namely a $U(2)$ global symmetry which will be spontaneously broken, with a source for the time-like component of the abelian $U(1)$ current, i.e. a chemical potential, which explicitly breaks boost invariance (but preserves $U(2)$).
This model has been previously considered in \cite{Amado:2013xya}, 
where the system was set at finite temperature and numerical techniques were essentially employed to analyze it (see also \cite{Filev:2009xp} for a different holographic model). We will rather consider the model at zero temperature, in analogy with the original field theory model presented in \cite{Miransky:2001tw,Schafer:2001bq}, and perform a purely boundary analysis through analytical techniques.

For this aim we will employ the technique of holographic renormalization \cite{Bianchi:2001de,Bianchi:2001kw} in order to deduce the presence of such Goldstone bosons purely by symmetry arguments. In practical terms, the Ward identities for broken symmetries \cite{Goldstone:1962es} establish the existence of massless modes generated by the broken currents. Moreover in the non-relativistic case, when commutators of broken generators have VEVs, it can be shown that some of these massless modes must have quadratic dispersion relations \cite{Brauner:2007uw}. By carefully performing the procedure of obtaining the renormalized action of the holographic model, we will show that such Ward identities are exactly reproduced. Unfortunately, this is all we can extract from a purely boundary analysis. This allows us to determine the existence of different type of Goldstone bosons and to establish if their dispersion relation is either linear ($\omega=c_{s}k$, with $c_{s}$ ``sound'' velocity) or quadratic ($\omega=\frac{k^2}{M}$, with $M$ a parameter with the dimensions of a mass), but the exact values of the constant factors $c_{s}$ and $M$ remain undetermined. In order to explicitly derive the dispersion relations and quantitatively determine these factors, an analysis of bulk fluctuations is required, similarly as it is necessary in order to determine the gap towards the massive excitations.

The note is structured as follows. We first present the holographic model and derive the renormalized action. Then by paying attention to gauge invariance, we extract the relevant correlators and find that they satisfy the expected Ward identities. We end by commenting on how to go beyond. In the appendices, we consider an alternative gauge fixing of the holographic model, and we show in the field theory model of \cite{Miransky:2001tw,Schafer:2001bq} how the non-local parts of the correlators, which can be explicitly computed, conspire to satisfy the same Ward identities, while yielding the expected dispersion relations.

\section{A holographic model for non-relativistic symmetry breaking}

We now outline the model discussed in \cite{Amado:2013xya}. This is a typically strongly coupled theory represented by its  holographic dual. 
We will assume that the theory has a non-trivial UV conformal fixed point. In this UV CFT, we will focus on the conserved currents $J_i^a$ which form a $U(2)$ algebra, and on a relevant operator $\CO_\Psi$ which is a doublet of $U(2)$. In the holographic dual, this means that we need to have a bulk theory in an asymptotically AdS spacetime, which includes dynamical $U(2)$ gauge fields and a complex doublet scalar with negative squared mass. Since we are not interested in computations involving the stress energy tensor, we set gravity in the bulk not to be dynamical. We thus consider the following bulk action, where for simplicity we choose four bulk dimensions (and hence three dimensions for the boundary theory):\footnote{We argue later that the choice of higher dimensions should not qualitatively modify our results.}
\begin{equation}
S_\mathrm{bulk}= \int d^4x \sqrt{-g}\left\{  -\frac{1}{4}F_{\mu\nu}^a F^{a\mu\nu} -D_\mu \Psi^\dagger D^\mu \Psi -m^2 \Psi^\dagger\Psi  \right\}\ ,
\label{Sbulk}
\end{equation}
where $F_{\mu\nu}^a=\partial_\mu A_\nu^a-\partial_\nu A_\mu^a+f^{abc}A_\mu^bA_\nu^c$ and $D_\mu \Psi = \partial_\mu \Psi -i A_\mu^a T^a \Psi$. The field $\Psi\equiv\left(\begin{smallmatrix} \lambda \\ \phi \end{smallmatrix}\right)$ is a complex scalar doublet, while $T^a$, with $a\!=\!\{0,1,2,3\}$, are the generators of $U(2)$ in the fundamental representation, namely $T^0=\frac12 \mathbbm{1}$ and $T^a=\frac12 \tau^a$ otherwise, with $\tau^a$ the Pauli matrices,
and they satisfy $[T^a,T^b]\!=\!if^{abc}T^c$, with $f^{abc}\!=\!0$ if any of the indices is 0 or $f^{abc}\!\equiv\!\epsilon^{abc}$ otherwise.

The metric is fixed to be the AdS one, defined by
\begin{equation}
ds^2=\frac{1}{z^2}(\eta_{ij}dx_idx_j +dz^2)\ ,
\end{equation}
where $\eta_{ij}$ is the mostly-plus Minkowski metric. Note that since this is pure AdS, on the field theory side it corresponds to considering zero temperature, in contrast with the analysis of \cite{Amado:2013xya} where a non-zero temperature was always required.

We will not consider any back-reaction on the metric. This is motivated by the fact that we will in any case restrict our attention to the near-boundary region, where back-reaction effects can be shown to be sub-leading and actually completely irrelevant to our considerations.

Let us first consider the field configuration defining the background, and thus the vacuum of the dual theory. First of all, as anticipated, the theory is non-relativistic because boost invariance is broken by the presence of a chemical potential, $\CL_\mathrm{QFT}\supset \mu J^0_t$.
This means that there should be a non-trivial source for $ J^0_t$, namely $A_t^0$ should have a leading mode turned on. This source breaks Lorentz invariance, but preserves all of $U(2)$ since the $U(1)$ generator commutes with all the algebra. In the vacuum of the theory, we expect that non-trivial dynamics generates VEVs for the operator $\CO_\Psi$, thus breaking $U(2)$ to $U(1)$. For simplicity we will take this $U(1)$ to be the one generated by $T^0+T^3$. The unbroken symmetries allow for VEVs to be generated also for $J_t^0$ and $J_t^3$. The latter is crucial for obtaining VEVs for commutators of charges, and thus for the appearance of Goldstone bosons with quadratic dispersion relations. In the bulk, we will thus have sub-leading modes for the profiles of $A_t^0$, $A_t^3$ and for $\Psi$ in its bottom component. 

From now on, we fix the dimension of $\CO_\Psi$ to be two, which implies $m^2=-2$ (in units of the AdS radius). The background profiles are thus the following: 
\begin{equation}
\left.\phi\,\rule[-2.6mm]{0.2mm}{6mm}\right._{B} = \bar\phi z^2, \quad \left.A_t^0\,\rule[-2.6mm]{0.2mm}{6mm}\right._{B} =\mu + \bar A^0 z\ , \quad \left.A_t^3\,\rule[-2.6mm]{0.2mm}{6mm}\right._{B} = \bar A^3 z\ ,
\end{equation}
where we take $\bar{\phi}$, $\mu$, $\bar A^0$ and $\bar A^3$ to be real constants, and all the other fields vanish. These profiles satisfy the free equations of motion. As with the metric, the back-reaction of the above profiles on each other can be safely neglected.\footnote{The backreaction would intervene at $\CO (z^4)$ in $\phi$ and at $\CO (z^2)$ in $A^{0,3}_t$, hence at orders that do not contribute to the rest of our computations.}
Strictly speaking, without backreaction the leading and subleading modes can be chosen independently. In particular $\bar A^{0,3}$ do not depend on $\mu$, which could be even set to zero. However this would mean that some physical mechanism should generate spontaneously a Lorentz breaking VEV such as $\langle J_t^3\rangle$. In the following we will rather assume that $\mu\neq 0$ even if it is not technically necessary.

We now proceed to fluctuate the fields over the background. The aim is to obtain the on-shell action up to quadratic order, since we will be interested in two- and one-point functions. At this order, the on-shell action reduces to a boundary term. In the following, we fix as usual the axial gauge $A_z^a=0$. The linearized equations of motion are the following (we use the notation $i=\{t,x,y\}$ and $u=\{x,y\}$, and we label the fluctuations above the background in the same way as the field itself). The constraint coming from the variation with respect to $A_z^a$ reads:
\begin{align}
z\partial_z\partial_i A_i^a & -z\bar A^3 f^{3ab} \left(A_t^b -z \partial_zA_t^b\right) -i\bar{\phi}(T^a)_{21} \big(2\lambda-z\partial_z \lambda\big) \nn
&+i\bar{\phi}(T^a)_{12} \left(2\lambda^\dagger-z\partial_z \lambda^\dagger\right)
-i\bar{\phi}(T^a)_{22} \left(2(\phi-\phi^\dagger)-z\partial_z(\phi-\phi^\dagger)\right) =0 \ .\label{eomAz}
 \end{align} 
The equations of motion for the vector field fluctuations, separated in temporal and spatial components, are
\begin{align}
z^2\partial_z^2 A_t^a+&z^2\partial_u^2A_t^a-z^2\partial_t\partial_uA_u^a +z^3f^{ab3}\bar A^3\partial_u A_u^b \nn
&+iz^2\bar{\phi} \left[(T^a)_{12}\partial_t\lambda^\dagger -(T^a)_{21}\partial_t\lambda-(T^a)_{22}\partial_t(\phi-\phi^\dagger)\right] \nn
&-z^2(\mu+\bar A^0z)\bar \phi \left[(T^a)_{21} \lambda+ (T^a)_{12}\lambda^\dagger +  (T^a)_{22} (\phi+\phi^\dagger)\right]  \nn
&-z^3 \bar A^3 \bar \phi \left[\{T^a,T^3\}_{21} \lambda+ \{T^a,T^3\}_{12}\lambda^\dagger +\{T^a,T^3\}_{22} (\phi+\phi^\dagger)\right] \nn
&-z^4 \bar{\phi}^2 A_t^b \{T^a,T^b\}_{22}=0\ , \phantom{\frac{a}{\big]}} \\
z^2\partial_z^2 A_u^a +&z^2\partial_i^2A_u^a -z^2\partial_u\partial_iA_i^a +2z^3f^{ab3}\bar A^3\partial_t A_u^b -z^3f^{ab3}\bar A^3\partial_u A_t^b -z^4 {(\bar A^3)}^2 f^{ab3}f^{bc3}A_u^c  \nn
&+iz^2\bar{\phi} \left[(T^a)_{12}\partial_u\lambda^\dagger-(T^a)_{21}\partial_u\lambda-(T^a)_{22}\partial_u(\phi-\phi^\dagger)\right] \nn
&-z^4 \bar{\phi}^2 A_u^b \{T^a,T^b\}_{22}=0\ .
\end{align}
Finally, the equations for the scalar fluctuations are
\begin{align}
z^2\partial_z^2\lambda-& 2z\partial_z\lambda +z^2\partial_i^2\lambda+2\lambda-iz^4\bar{\phi} (T^a)_{12}\partial_i A_i^a \nn 
&+ iz^2(\mu+\bar A^0z+\bar A^3z)\partial_t\lambda +\frac14 z^2(\mu+\bar A^0z+\bar A^3z)^2\lambda \nn 
& +z^4 (\mu+\bar A^0z)\bar\phi(T^a)_{12} A_t^a =0\ , \phantom{\frac{a}{\big]}} \\
z^2\partial_z^2\phi-& 2z\partial_z\phi +z^2\partial_i^2\phi+2\phi-iz^4\bar{\phi} (T^a)_{22}\partial_i A_i^a\nn & + iz^2(\mu+\bar A^0z-\bar A^3z)\partial_t\phi 
+ \frac14 z^2(\mu+\bar A^0z-\bar A^3z)^2\phi \nn & +z^4 (\mu+\bar A^0z)\bar\phi(T^a)_{22} A_t^a +z^5 \bar A^3\bar\phi\{T^a,T^3\}_{22} A_t^a=0\ .
\end{align}
The last two equations should be supplemented by their complex conjugates. 

The on-shell action  is obtained expanding \eqref{Sbulk} up to quadratic order in the fluctuations, and then substituting the equations of motion, possibly integrating by parts. 

At the regularizing surface $z=\epsilon$, we obtain
\begin{align}
S_\mathrm{reg}=-\int_{z=\epsilon} d^3x &\left\{    \bar A^0 A_t^0+\bar A^3 A_t^3-\frac{2}{z}\bar{\phi}(\phi+\phi^\dagger) +\frac12 A_t^a \partial_z A_t^a -  \frac12 A_u^a \partial_z A_u^a\right. \nn
&\left.  \qquad-  \frac1 {2z^2} \left(\lambda^\dagger \partial_z \lambda
+\lambda \partial_z \lambda^\dagger+\phi^\dagger \partial_z \phi +
\phi \partial_z \phi^\dagger \right) \right\} \ . \label{sreg}
\end{align}
At this point, we note that the quadratic terms are exactly the same that would arise in a configuration with vanishing backgrounds. The presence of non-trivial backgrounds must then show up when expanding the fluctuations near the boundary as powers of $z$. There is however one more substitution that we can make, that makes the dependence on the background manifest even before expanding the fluctuations. We can indeed also substitute the equation \eqref{eomAz} which only has first order derivatives in $z$. However in order to perform the substitution we also have to split the vector into its irreducible components. We choose here to split into transverse and longitudinal parts with respect to the spatial coordinates $u$. Thus we keep $A_t^a$ as the temporal component, while we split\footnote{In Appendix A we discuss a different splitting, where the transverse and longitudinal parts are taken with respect to the spacetime coordinates $i$.}
\begin{equation}
A_u^a = A_u^{Ta}+ \partial_u A^{La}, \qquad  \partial_u A_u^{Ta}=0\ .
\label{split1}
\end{equation}
We eventually arrive at 
\begin{align}
S_\mathrm{reg}=-\int_{z=\epsilon} d^3x &\left\{    \bar A^0 A_t^0+\bar A^3 A_t^3-\frac{2}{z}\bar{\phi}(\phi+\phi^\dagger) +\frac12 A_t^a \partial_z A_t^a -  \frac12 A_u^{Ta} \partial_z A_u^{Ta}\right. \nn
&\qquad+\frac12 A^{La} \partial_z\partial_tA_t^a +\frac12\bar A^3 f^{ab3} A^{La} (A_t^b - z\partial_zA_t^b) \nn
&\qquad +\frac{i}{2z}  \bar \phi A^{La} \left[ (T^a)_{21}( 2\lambda-z\partial_z\lambda) -(T^a)_{12}( 2\lambda^\dagger-z\partial_z\lambda^\dagger) \right. \nn
&\qquad \qquad\qquad\qquad\left. +(T^a)_{22}(2\phi-2\phi^\dagger-z\partial_z\phi +z\partial_z\phi^\dagger)\right]\nn
&\left.  \qquad-  \frac1 {2z^2} \left(\lambda^\dagger \partial_z \lambda
+\lambda \partial_z \lambda^\dagger+\phi^\dagger \partial_z \phi +
\phi \partial_z \phi^\dagger \right) \right\} \ .
\end{align}
We now consider the near-boundary expansion of the fluctuating fields:
\begin{align}
A_t^a &= A_{t(0)}^a+ A_{t(1)}^a z +\dots \nn
A_u^{Ta} &= A_{u(0)}^{Ta}+ A_{u(1)}^{Ta} z +\dots \nn
A^{La} &= A_{(0)}^{La}+ A_{(1)}^{La} z +\dots \nn
\lambda&= \lambda_{(0)} z +  \lambda_{(1)} z^2 + \dots \nn
\phi&= \phi_{(0)} z +  \phi_{(1)} z^2 + \dots
\label{expansions}
\end{align}
Note that all the modes with ${}_{(0)}$ subscript are the ones which are usually considered as sources, while all the ones with ${}_{(1)}$ subscript coincide with the independent sub-leading modes, to be identified with the VEVs of the fluctuations. The latter occur at the next-to-leading order in the $z$ expansion because of the fact that we are considering conserved currents and an operator of dimension 2 in a three-dimensional boundary theory.

The divergent terms that one finds in $S_\mathrm{reg}$ are taken care of by adding counter-terms, which are independent of the presence of the profile. Again, due to the involved dimensions, no finite counter-terms arise and hence no scheme dependence.\footnote{If we had considered an operator of dimension 3 in a four-dimensional boundary theory, for instance, we would have logarithmical terms in the expansions of the fluctuations which would lead to a logarithmical divergence in the regularized action. The renormalization of the logarithmical terms carries along finite counter-terms, that would be scheme-dependent. However, as it should be clear in the following, this ambiguity would not affect the terms we are interested in.}
Eventually, the renormalized action is the following:
\begin{align}
S_\mathrm{ren}=-\int d^3x &\left\{    \bar A^0 A_{t(0)}^0+\bar A^3 A_{t(0)}^3-2\bar{\phi}(\phi_{(0)}+\phi_{(0)}^\dagger) +\frac12 A_{t(0)}^a  A_{t(1)}^a -  \frac12 A_{u(0)}^{Ta}  A_{u(1)}^{Ta}\right. \nn
&\qquad-\frac12\partial_t A^{La}_{(0)} A_{t(1)}^a +\frac12\bar A^3 f^{ab3} A^{La}_{(0)} A_{t(0)}^b  \nn
&\qquad +\frac{i}{2}  \bar \phi A^{La}_{(0)} \left[ (T^a)_{21} \lambda_{(0)} -(T^a)_{12}\lambda_{(0)}^\dagger  +(T^a)_{22}(\phi_{(0)}-\phi_{(0)}^\dagger)\right]\nn
&\left.  \qquad-  \frac1 {2} \left(\lambda_{(0)}^\dagger  \lambda_{(1)}
+\lambda_{(0)} \lambda_{(1)}^\dagger+\phi_{(0)}^\dagger  \phi_{(1)} +
\phi_{(0)}  \phi_{(1)}^\dagger \right) \right\} \ . \label{Sren}
\end{align}
The first three terms of the above functional, linear in the fluctuations, just give us the VEVs of the associated operators $J_t^0$, $J_t^3$ and $\mathrm{Re} \CO_\phi$.
The other terms, quadratic in the fluctuations, contain in principle all the information about two-point functions. However, this information is encoded in the way the sub-leading modes with ${}_{(1)}$ subscript depend on the sources 
with ${}_{(0)}$ subscript. This dependence is fixed through the bulk boundary conditions (generically, asking regularity) and is typically non-local, since it requires solving the equations for the fluctuations inside the bulk. We now turn to see what can be extracted from $S_\mathrm{ren}$ without solving the bulk equations of motion.

\section{Ward identities from the renormalized action}

A brief inspection of $S_\mathrm{ren}$ in \eqref{Sren} shows that there are two kinds of quadratic terms: those which are bilinears of a source and a VEV of the fluctuations and those which involve only sources. All the latter ones are proportional to a VEV of the non-trivial background we have chosen, thus they would not be present in a model with trivial profiles and only the former ones would survive. Nonetheless it turns out that terms of the second kind are also hidden into terms of the first kind, because of gauge invariance. 

Consider indeed the bulk gauge invariance of the action \eqref{Sbulk}, for which $\delta \Psi= i\alpha \Psi$ and $\delta A_\mu = \partial_\mu \alpha +i[\alpha,A_\mu]$. Because of the gauge fixing $A_z^a=0$, the gauge parameter does not depend on $z$. It is then easy to see how the residual gauge transformations act on each mode in the expansions
\eqref{expansions}. Recalling further that we are only interested in the quadratic part of the action, we neglect terms in the gauge variations which are bilinear in the gauge parameter and the mode of the fluctuation.
We are then left with:
\begin{align}
&\delta \lambda_{(0)} = 0 \ , \qquad \quad \delta \lambda_{(1)} = i \alpha^a (T^a)_{12} \bar{\phi} \ , \nn
&\delta \phi_{(0)} = 0 \ , \qquad\quad  \delta \phi_{(1)} = i \alpha^a (T^a)_{22} \bar{\phi} \ , \nn
&\delta A_{t(0)}^a = \partial_t \alpha^a\ , \qquad \delta A_{t(1)}^a = -f^{ab3} \alpha^b \bar A^3 \ ,\nn
&\delta A_{u(0)}^{Ta} = 0 \ , \qquad \quad \delta A_{u(1)}^{Ta} = 0 \ , \nn
&\delta A_{(0)}^{La} = \alpha^a \ , \qquad \quad \delta A_{(1)}^{La} = 0\ .
\end{align}
After solving the bulk equations of motion for the fluctuations, the sub-leading modes $\Phi_{(1)}$ will be expressed in terms of non-local functions of the sources 
$\Phi_{(0)}$. However, in order to solve the equations, we would have to impose boundary conditions in the bulk and, to preserve the gauge symmetry, we should take care of imposing them on gauge-invariant combinations of the fields. So we have to consider only gauge-invariant combinations of both the sources and the sub-leading modes, as for instance
\begin{equation}
\lambda_{(1)}-i\bar \phi (T^a)_{12}  A_{(0)}^{La}\ , \qquad 
\phi_{(1)} -i \bar \phi (T^a)_{22} A_{(0)}^{La}\ , \qquad 
 A_{t(1)}^a +\bar A^3 f^{ab3} A_{(0)}^{Lb}\ ,
\end{equation}
for the VEVs and
\begin{equation}
A_{t(0)}^a - \partial_tA_{(0)}^{La}\ ,
\end{equation}
for the sources.

We are now allowed to assume the general relations between VEVs and sources:
\begin{align}
\lambda_{(1)}&=i\bar \phi (T^a)_{12}  A_{(0)}^{La} +f(\partial) \lambda_{(0)} + \tilde f^a (\partial) (A_{t(0)}^a - \partial_tA_{(0)}^{La}) \ , \nn
\phi_{(1)}& =i \bar \phi (T^a)_{22} A_{(0)}^{La} + g(\partial)\phi_{(0)} + \tilde  g^a(\partial) (A_{t(0)}^a - \partial_tA_{(0)}^{La}) \ , \nn 
A_{t(1)}^a &= -\bar A^3 f^{ab3} A_{(0)}^{Lb} + h^{ab}(\partial)  (A_{t(0)}^b - \partial_tA_{(0)}^{Lb})\nn
&\qquad +k^a(\partial) \lambda_{(0)}
+l^a(\partial)\phi_{(0)} +k^a(\partial)^* \lambda_{(0)}^\dagger
+l^a(\partial)^*\phi_{(0)}^\dagger\ ,\nn
A_{u(1)}^{Ta} &= m^{ab}(\partial) A_{u(0)}^{Tb}\ ,\label{vevsources}
\end{align}
where all the functions of $\partial$ collectively indicate expressions that are typically non-local in space and/or time derivatives, and that cannot be determined without solving the equations in the bulk. 
However, we will see that in some combinations of the correlators the dependence on these unknown functions drops out. 

We can now rewrite $S_\mathrm{ren}$ eliminating all the VEVs, using the expressions above:
\begin{align}
S_\mathrm{ren}=-\int d^3x &\left\{    \bar A^0 A_{t(0)}^0+\bar A^3 A_{t(0)}^3-2\bar{\phi}(\phi_{(0)}+\phi_{(0)}^\dagger) -  \frac12 A_{u(0)}^{Ta} m^{ab}(\partial) A_{u(0)}^{Tb}\right. \nn
&\qquad+\frac12 (A_{t(0)}^a   -\partial_t A^{La}_{(0)}) h^{ab}(\partial)  (A_{t(0)}^b   -\partial_t A^{Lb}_{(0)}) \nn
&\qquad -\frac12\bar A^3 f^{ab3} (2A_{t(0)}^a -\partial_t A^{La}_{(0)})A^{Lb}_{(0)}  \nn
&\qquad +i  \bar \phi A^{La}_{(0)} \left[ (T^a)_{21} \lambda_{(0)} -(T^a)_{12}\lambda_{(0)}^\dagger  +(T^a)_{22}(\phi_{(0)}-\phi_{(0)}^\dagger)\right]\nn
&\qquad -\frac12 \lambda_{(0)}^\dagger (f(\partial)+f(\partial)^*) \lambda_{(0)} -\frac12  \phi_{(0)}^\dagger (g(\partial)+g(\partial)^*)\phi_{(0)} \nn
& \qquad-\frac12 \left[\lambda_{(0)}^\dagger (\tilde f^a (\partial)+k^a(\partial)^*)+\lambda_{(0)} (\tilde f^a (\partial)^*+k^a(\partial)) \right] (A_{t(0)}^a - \partial_tA_{(0)}^{La}) \nn
&\qquad\left. -\frac12 \left[\phi_{(0)}^\dagger ((\tilde g^a (\partial)+l^a(\partial)^*)+\phi_{(0)} (\tilde g^a (\partial)^*+l^a(\partial))\right] (A_{t(0)}^a - \partial_tA_{(0)}^{La})\right\} \ .\label{srenfinal}
\end{align}
This is the generating functional for the one- and two-point functions in our theory.\footnote{Note that the scheme-ambiguity which would arise in higher dimensions would be contained in the possibility to redefine the non-local functions in the above expression.} The precise relations between sources of operators in the boundary theory and modes of bulk fluctuations are the following. For the scalar operators we have
\begin{equation}
\int_{\partial AdS} d^3x \left( \lambda_{(0)} \CO_\lambda + \phi_{(0)} \CO_\phi + c.c.\right)\ ,
\end{equation}
while for the currents
\begin{equation}
\int_{\partial AdS} d^3x \left( A_{t(0)}^a J_t^a - A_{u(0)}^{Ta}J_u^{Ta} +A_{(0)}^{La} \partial_u J_u^a \right)\ ,
\end{equation}
so that $A_{u(0)}^{Ta}$ sources the purely transverse part of $J_u^a$ while $A_{(0)}^{La}$ its longitudinal piece.

Some two-point functions will be entirely determined by their non-local part, for instance those with two transverse currents or two scalar operators, and we will have nothing to say about them since we do not solve the bulk equations. On the other hand, we see from the final expression of our generating functional $S_\mathrm{ren}$ that some other two-point functions might be directly determined by our analysis. It should be the case for two-point functions involving the temporal and longitudinal components of the currents, both among themselves or mixed with scalar operators. Indeed, local constant terms involving the sources of these operators appear in \eqref{srenfinal}.

Let us list here a number of such correlators:
\begin{align}
\langle J_t^a (x) J_t^b(y) \rangle  &= -i \frac{\delta^2 S_\mathrm{ren}}{\delta A_{t(0)}^a(x) \delta  A_{t(0)}^b} (y)= i h^{ab}(\partial) \delta^3(x-y)\ ,\label{corrjtjt}\\
\langle J_t^a(x) \partial_u J_u^b (y)\rangle  &= -i \frac{\delta^2 S_\mathrm{ren}}{\delta A_{t(0)}^a (x)\delta  A_{(0)}^{Lb}(y)} = -i\left[ h^{ab}(\partial)\partial_t +\bar A^3 f^{ab3}\right]\delta^3(x-y)\ ,\\
\langle \partial_u J_u^a(x) \partial_v J_v^b (y)\rangle  &= -i \frac{\delta^2 S_\mathrm{ren}}{\delta A_{(0)}^{La} (x)\delta  A_{(0)}^{Lb}(y)} = -i\left[h^{ab}(\partial)\partial_t^2+\bar A^3 f^{ab3}\partial_t\right] \delta^3(x-y)\ ,\\
\langle J_t^a (x) \CO_\lambda (y)\rangle &= -i \frac{\delta^2 S_\mathrm{ren}}{\delta A_{t(0)}^a(x) \delta \lambda_{(0)}(y)} = 
-\frac{i}{2} \left(\tilde f^a (\partial)^*+k^a(\partial)\right) \delta^3(x-y)\ ,\\
\langle \partial_u J_u^a(x) \CO_\lambda (y)\rangle &= -i \frac{\delta^2 S_\mathrm{ren}}{\delta A_{(0)}^{La}(x) \delta \lambda_{(0)}(y)} = i\left[
-\frac{1}{2} \left(\tilde f^a (\partial)^*+k^a(\partial)\right)\partial_t +i\bar \phi  (T^a)_{21} \right] \delta^3(x-y)\ ,\\
\langle J_t^a (x) \CO_\phi (y)\rangle &= -i \frac{\delta^2 S_\mathrm{ren}}{\delta A_{t(0)}^a(x) \delta \phi_{(0)}(y)} = 
-\frac{i}{2} \,\Big(\tilde g^a (\partial)^*+l^a(\partial)\Big)\, \delta^3(x-y)\ ,\\
\langle \partial_u J_u^a(x) \CO_\phi (y)\rangle &= -i \frac{\delta^2 S_\mathrm{ren}}{\delta A_{(0)}^{La}(x) \delta \phi_{(0)}(y)} = i\left[-\frac{1}{2} \Big(\tilde g^a (\partial)^*+l^a(\partial)\Big)\partial_t +i\bar \phi  (T^a)_{22} \right] \delta^3(x-y)\ .\label{corrjuop}
\end{align}
We thus immediately see that some combinations are given entirely by the constant terms, or trivially vanish:
\begin{align}
-
\langle\partial_t J_t^a (x) J_t^b(y) \rangle + \langle \partial_u J_u^a(x)J_t^b (y) \rangle &= - i \bar A^3 f^{ab3}\delta^3(x-y)\ ,\label{witemp}\\
-
\langle \partial_tJ_t^a(x) \partial_v J_v^b (y)\rangle 
+
\langle \partial_u J_u^a(x) \partial_v J_v^b (y)\rangle  &= 0\ ,\label{wilong}\\
-
\langle \partial_tJ_t^a (x) \CO_\lambda (y)\rangle
+
\langle \partial_u J_u^a(x) \CO_\lambda (y)\rangle &= 
-
\bar \phi  (T^a)_{21}  \delta^3(x-y)\ ,\label{wilambda}\\
-
\langle \partial_tJ_t^a (x) \CO_\phi (y)\rangle
+
\langle \partial_u J_u^a(x) \CO_\phi (y)\rangle &= 
-
\bar \phi  (T^a)_{22}  \delta^3(x-y)\ .\label{wiphi}
\end{align}
These are of course nothing else than the Ward identities relating the two-point functions of currents associated to broken generators, to the VEVs of the operators that break the symmetry. In particular, the relations \eqref{wilambda} and \eqref{wiphi} are the usual identities relating the two-point function of the divergence of a conserved current and a scalar operator to the  symmetry breaking VEV of the operator. This kind of Ward identity has been already realized holographically, see e.g.~\cite{Bianchi:2001de,Bianchi:2001kw} (and more recently also \cite{Argurio:2014rja}). 


Of more interest is the identity \eqref{witemp}, which is non-trivial due to the fact that we allow the temporal component of a current ($J_i^3$ here) to have a non-trivial VEV. We assume that this Lorentz symmetry breaking VEV is permitted by the presence of a chemical potential, though in our holographic set-up this is not technically necessary (indeed $\mu$ does not explicitly appear anywhere in the above expressions). In addition, note that the identity \eqref{wilong} is consistently trivial since the spatial (longitudinal) components of the same current cannot get a VEV, because that would violate the invariance under spatial rotations.

The above Ward identities imply the presence of Goldstone bosons, i.e.~of massless modes in the spectrum. More precisely, we see that in order to satisfy the identities \eqref{witemp}--\eqref{wiphi}, the Fourier transformed correlators $\langle J_t^a  J_t^b \rangle(\omega,k)$, $\langle J_t^a  \CO_\lambda \rangle(\omega,k)$ and the similar ones must be singular when the energy $\omega$ and the momentum $k_u$ go to zero. Indeed, in Fourier space 
\eqref{witemp} reads
\begin{equation}
-i\omega \langle J_t^a  J_t^b \rangle+ik_u \langle  J_u^aJ_t^b  \rangle= -i \bar A^3 f^{ab3}\ .\label{wift}
\end{equation}
We thus deduce the presence of massless poles in all of these correlators. 
%
%
This additional Ward identity, which defines \cite{Watanabe:2012hr} type B Goldstone bosons, requires the dispersion relation to be quadratic \cite{Brauner:2007uw}, with the following argument.

In the Lorentz invariant case, the dispersion relation of a Goldstone boson is trivially determined from a Ward identity similar to \eqref{wilambda} and \eqref{wiphi}, giving
\begin{equation}
\langle J_i \, \CO \rangle \propto \frac{k_i}{\omega^2-k_u^2}\ .
\end{equation}
In lack of Lorentz invariance, we have to consider two different situations, one in which time-reversal invariance is preserved and one in which it is broken. When it is preserved, for small values of $\omega$ and $k_u$ we can admit:
\begin{equation}
 \langle J_t \, \CO \rangle \simeq \frac{\tilde T\omega}{\omega^2-c_s^2 k^2}\ , \qquad  \langle J_u \, \CO \rangle \simeq \frac{\tilde{U} k_u}{\omega^2-c_s^2 k^2}\ ,
  \label{tcorr} 
\end{equation}
while when it is broken, we can have 
\begin{equation}
\langle J_t \, \CO \rangle \simeq  \frac{\bar T}{M\omega- k^2}\ , \qquad  \langle J_u \, \CO \rangle \simeq  \frac{\tilde{U}k_u}{M\omega- k^2}\ ,
\end{equation}
where $\bar T$, $\tilde T$, $\tilde U$, $c_s$ and $M$ are constants. 
Note now that \eqref{wift} breaks time-reversal invariance for the currents $J^1_i$ and $J^2_i$, and actually requires $\bar T\!\neq\!0$ as it was proved in \cite{Brauner:2007uw}. This implies that \eqref{witemp} and \eqref{wilambda} lead to  quadratic dispersion relations,  $\omega\simeq\frac{k^2}{M}$ with $M\equiv{\bar T}/{\bar \phi}$. On the other hand for the Goldstone boson contributing to \eqref{wiphi}, which is of type~A \cite{Watanabe:2012hr}, the time-reversal invariance is still preserved and the dispersion relation is linear (assuming $\tilde{U}$ has a finite limit for vanishing momentum), but with velocity depending on the ratio $\tilde{U}/\tilde{T}\equiv c_s^2$.

Without specifying the unknown non-local functions $h^{ab}(\partial)$, $f^a(\partial)$, etc., we cannot go further and, for instance, find the exact expression for $\bar T$, $\tilde{T}$, $\tilde{U}$ for all the massless (and light) excitations. In the present model this is of course in principle possible (see \cite{Amado:2013xya} for a finite temperature analysis), but would imply solving the equations of motion for the fluctuations in the bulk. This in turn would necessitate to find the back-reacted geometry. The point in the present note was to exploit to its limits the technique of holographic renormalization, i.e.~to extract the maximal information on the system purely from boundary considerations. Perhaps not unexpectedly, we found exactly the same information that can be gathered from Ward identities that apply to the system. 

In Appendix B we compute the same correlators in the field theoretical toy model of \cite{Miransky:2001tw,Schafer:2001bq}, where both the local and the non-local parts of the correlators can be made explicit.

\section{Conclusions and perspectives}

We set out to compute two-point correlators through holographic renormalization, in a strongly coupled field theory model with a global $U(2)$ symmetry, where a chemical potential breaks explicitly boost invariance. We  studied the consequences of the spontaneous breaking of $U(2)$ to $U(1)$. Because of the broken Lorentz symmetry, not only scalar operators are allowed to acquire symmetry breaking VEVs, but also temporal components of conserved currents. 

We have obtained the general form of the renormalized action up to quadratic order, that is the generating functional for one- and two-point correlation functions. In this expression, we have kept implicit the non-local functions that are established by imposing regularity conditions of the fluctuations in the deep bulk. Nevertheless, we were able to extract information on the light spectrum by showing that some linear combinations of the correlators are completely determined by local terms, and  actually realize holographically the Ward identities associated to the broken symmetries. 

We have thus established that the system must have Goldstone excitations associated to the three broken generators. However, since we have also shown that $J_t^3$, and thus $Q^3$, has a VEV, we observe that we are exactly in a situation where so-called type B Goldstone bosons \cite{Watanabe:2012hr} arise, i.e.~when the commutator of two charges has a non-vanishing VEV. Here, we have that $\langle [Q^1,Q^2]\rangle \neq 0$, so that only one massless excitation is associated to these two broken charges. Furthermore, we have given an argument for this massless excitation to have quadratic dispersion relation. The remaining broken generator $Q^0-Q^3$ is not involved in commutators with non-trivial VEVs, and hence gives rise to a type A Goldstone boson, which in this case is expected to have a linear dispersion relation, but with a velocity smaller than $c$ that would be determined by the explicit expressions of the non-local parts of the correlators. Moreover, we expect the type B Goldstone boson to be accompanied by an almost Goldstone boson, i.e.~a light mode whose mass is related to the coefficient of the quadratic dispersion relation of its partner \cite{Kapustin:2012cr}. 

We stress once more that in order to obtain quantitive results on all these dispersion relations, we have to compute the non-local functions that we have left unspecified in \eqref{vevsources}. The poles in these functions will give us the dispersion relations of the massless modes, together with all the rest of the massive spectrum. In order to do that, we would need first to have a background which is reliable down to the deep bulk. Performing the back-reaction, also on the metric, would then be necessary. One should be warned though that in the present zero-temperature set-up, that would most inevitably lead to a singular geometry. That should however not prevent us from imposing boundary conditions in the form of boundedness of the fluctuations. Any other stratagem to avoid the singularity would introduce a new scale to the problem, as for instance a finite temperature. That has been done in \cite{Amado:2013xya}, where however the back-reaction is not studied, thus limiting the analysis to situations where the temperature and the chemical potential are roughly at the same scale.

Considering finally generalizations, it is obviously rather straightforward to generalize our discussion to higher dimensions. It would be interesting to investigate, in a fully back-reacted model, at zero or non-zero temperature, also correlators involving the stress-energy tensor, and possibly in supersymmetric extensions of such models, whether also the dispersion relations of the Goldstino can be modified, extending the analysis of \cite{Argurio:2013uba,Argurio:2014uca}.

\section*{Acknowledgments}

We are grateful to Matteo Bertolini, Daniele Musso and Diego Redigolo for useful discussions and feedback on the manuscript, and Manuela Kulaxizi and Andrei Parnachev for interesting comments. This research  is supported in part by IISN-Belgium (convention 4.4503.15), by the ``Communaut\'e
Fran\c{c}aise de Belgique" through the ARC program and by a ``Mandat d'Impulsion Scientifique" of the F.R.S.-FNRS. R.A. is a Senior Research Associate of the Fonds de la Recherche Scientifique--F.N.R.S. (Belgium).
The work of A.Me.($\slashed\pounds$) is supported by the NWO Vidi grant.


\appendix

\section{Alternative splitting of the vector degrees of freedom}
In this Appendix we perform, for completeness, a different splitting of the vector degrees of freedom. Instead of \eqref{split1}, we will split the vector into transverse and longitudinal parts with respect to all spacetime components:
\begin{equation}
A_i^a = a_i^{Ta}+ \partial_i a^{La}, \qquad  \partial_i a_i^{Ta}=0\ .
\label{split2}
\end{equation}
The regularized action \eqref{sreg} then reads, after the substitutions
\begin{align}
S_\mathrm{reg}=-\int_{z=\epsilon} d^3x &\left\{    \bar A^0 a_t^{T0}+\bar A^3 a_t^{T3}-\frac{2}{z}\bar{\phi}(\phi+\phi^\dagger)  -  \frac12 a_i^{Ta} \partial_z a_i^{Ta}\right. \nn
&\qquad +\frac12\bar A^3 f^{ab3} a^{La} (a_t^{Tb} - z\partial_za_t^{Tb}+\partial_t a^{Lb} - z\partial_z\partial_t a^{Lb}) \nn
&\qquad +\frac{i}{2z}  \bar \phi\ a^{La} \left[ (T^a)_{21}( 2\lambda-z\partial_z\lambda) -(T^a)_{12}( 2\lambda^\dagger-z\partial_z\lambda^\dagger) \right. \nn
&\qquad \qquad\qquad\qquad\left. +(T^a)_{22}(2\phi-2\phi^\dagger-z\partial_z\phi +z\partial_z\phi^\dagger)\right]\nn
&\left.  \qquad-  \frac1 {2z^2} \left(\lambda^\dagger \partial_z \lambda
+\lambda \partial_z \lambda^\dagger+\phi^\dagger \partial_z \phi +
\phi \partial_z \phi^\dagger \right) \right\} \ .
\end{align}
The near boundary expansions of the vector fields are now
\begin{align}
a_i^{Ta} &= a_{i(0)}^{Ta}+ a_{i(1)}^{Ta} z +\dots \nn
a^{La} &= a_{(0)}^{La}+ a_{(1)}^{La} z +\dots 
\label{expansionsbis}
\end{align}
and the renormalized action reads
\begin{align}
S_\mathrm{ren}=-\int d^3x &\left\{    \bar A^0 a_{t(0)}^{T0}+\bar A^3 a_{t(0)}^{T3}-2\bar{\phi}(\phi_{(0)}+\phi_{(0)}^\dagger) -  \frac12 a_{i(0)}^{Ta}  a_{i(1)}^{Ta}\right. \nn
&\qquad +\frac12\bar A^3 f^{ab3} a^{La}_{(0)} (a_{t(0)}^{Tb}+\partial_t a^{Lb}_{(0)})  \nn
&\qquad +\frac{i}{2}  \bar \phi\ a^{La}_{(0)} \left[ (T^a)_{21} \lambda_{(0)} -(T^a)_{12}\lambda_{(0)}^\dagger  +(T^a)_{22}(\phi_{(0)}-\phi_{(0)}^\dagger)\right]\nn
&\left.  \qquad-  \frac1 {2} \left(\lambda_{(0)}^\dagger  \lambda_{(1)}
+\lambda_{(0)} \lambda_{(1)}^\dagger+\phi_{(0)}^\dagger  \phi_{(1)} +
\phi_{(0)}  \phi_{(1)}^\dagger \right) \right\} \ .
\end{align}
The residual gauge transformations act on the transverse and longitudinal modes as:
\begin{align}
&\delta a_{i(0)}^{Ta} = 0 \ , \qquad \quad \delta a_{i(1)}^{Ta} = f^{ab3}  \bar A^3\left(\eta_{ti}-\frac{\partial_t\partial_i}{\partial_j^2} \right)\alpha^b\ , \nn
&\delta a_{(0)}^{La} = \alpha^a \ , \qquad \quad \delta a_{(1)}^{La} = f^{ab3}  \bar A^3\frac{\partial_t}{\partial_i^2}\alpha^b\ .
\end{align} 
Note that the gauge transformations of the VEV fluctuations are non-local with this choice of splitting. The gauge invariant combinations are 
\begin{equation}
\lambda_{(1)}-i\bar \phi (T^a)_{12}  a_{(0)}^{La}\ , \qquad 
\phi_{(1)} -i \bar \phi (T^a)_{22} a_{(0)}^{La}\ , \qquad 
 a_{i(1)}^{Ta} - \bar A^3 f^{ab3}\left(\eta_{ti}-\frac{\partial_t\partial_i}{\partial_j^2} \right) a_{(0)}^{Lb}\ .\label{gaugeinvbis}
\end{equation}
The expressions of the VEVs in terms of the sources are then
\begin{align}
\lambda_{(1)}&=i\bar \phi (T^a)_{12}  a_{(0)}^{La} +f(\partial) \lambda_{(0)} + \tilde f^a (\partial) a_{t(0)}^{Ta}  \ , \nn
\phi_{(1)}& =i \bar \phi (T^a)_{22} a_{(0)}^{La} + g(\partial)\phi_{(0)} + \tilde  g^a(\partial) a_{t(0)}^{Ta} \ , \nn 
a_{t(1)}^{Ta} &= - \bar A^3 f^{ab3} \left(1+\frac{\partial_t^2}{\partial_i^2} \right) a_{(0)}^{Lb} + h^{ab}(\partial) a_{t(0)}^{Tb} \nn
&\qquad +k^a(\partial) \lambda_{(0)}
+l^a(\partial)\phi_{(0)} +k^a(\partial)^* \lambda_{(0)}^\dagger
+l^a(\partial)^*\phi_{(0)}^\dagger\ ,\nn
a_{u(1)}^{Ta} &= -\bar A^3 f^{ab3} \frac{\partial_t\partial_u}{\partial_i^2}  a_{(0)}^{Lb} +m^{ab}(\partial) a_{u(0)}^{Tb}\ , \label{vevsourcesbis}
\end{align}
where we have used the same letters for the unknown non-local functions, which however will be possibly different from the ones in \eqref{vevsources}.

Eliminating all the VEVs, $S_\mathrm{ren}$ then reads:
\begin{align}
S_\mathrm{ren}=-\int d^3x &\left\{    \bar A^0 a_{t(0)}^{T0}+\bar A^3 a_{t(0)}^{T3}-2\bar{\phi}(\phi_{(0)}+\phi_{(0)}^\dagger) -  \frac12 a_{u(0)}^{Ta} m^{ab}(\partial) a_{u(0)}^{Tb}\right. \nn
&\qquad+\frac12 a_{t(0)}^{Ta}  h^{ab}(\partial)  a_{t(0)}^{Tb}   -\frac12\bar A^3 f^{ab3} (2a_{t(0)}^{Ta} +\partial_t a^{La}_{(0)})a^{Lb}_{(0)}  \nn
&\qquad +i  \bar \phi a^{La}_{(0)} \left[ (T^a)_{21} \lambda_{(0)} -(T^a)_{12}\lambda_{(0)}^\dagger  +(T^a)_{22}(\phi_{(0)}-\phi_{(0)}^\dagger)\right]\nn
&\qquad -\frac12 \lambda_{(0)}^\dagger (f(\partial)+f(\partial)^*) \lambda_{(0)} -\frac12  \phi_{(0)}^\dagger (g(\partial)+g(\partial)^*)\phi_{(0)} \nn
& \qquad-\frac12 \left[\lambda_{(0)}^\dagger (\tilde f^a (\partial)+k^a(\partial)^*)+\lambda_{(0)} (\tilde f^a (\partial)^*+k^a(\partial)) \right] a_{t(0)}^{Ta} \nn
&\qquad\left. -\frac12 \left[\phi_{(0)}^\dagger ((\tilde g^a (\partial)+l^a(\partial)^*)+\phi_{(0)} (\tilde g^a (\partial)^*+l^a(\partial))\right] a_{t(0)}^{Ta} \right\} \ .\label{srenfinalbis}
\end{align}
The coupling of the sources to the currents is now
\begin{equation}
\int_{\partial AdS} d^3x \left( a_{t(0)}^{Ta} J_t^{Ta} - a_{u(0)}^{Ta}J_u^{Ta} +a_{(0)}^{La} \partial_i J_i^a \right)\ .
\end{equation}
The Ward identities are then directly derived as
\begin{align}
\langle \partial_i J_i^a(x) J_t^{Tb} (y)\rangle &= - i \frac{\delta^2 S_\mathrm{ren}}{\delta a_{(0)}^{La}(x) \delta a_{t(0)}^{Tb}(y)} =-i \bar A^3 f^{ab3}\delta^3(x-y)\ ,\\
\langle \partial_i J_i^a(x) \CO_\lambda (y)\rangle &= -i \frac{\delta^2 S_\mathrm{ren}}{\delta a_{(0)}^{La}(x) \delta \lambda_{(0)}(y)} = - \bar \phi  (T^a)_{21}\delta^3(x-y)\ ,\\
\langle \partial_i J_i^a(x) \CO_\phi (y)\rangle &= -i \frac{\delta^2 S_\mathrm{ren}}{\delta a_{(0)}^{La}(x) \delta \phi_{(0)}(y)} =- \bar \phi  (T^a)_{22}\delta^3(x-y)\ ,
\end{align}
reproducing exactly the results of \eqref{witemp}--\eqref{wiphi}. 
The other correlators are purely in terms of the non-local parts, for instance
\begin{equation}
\langle J_t^{Ta} (x) \CO_\lambda (y)\rangle = -i \frac{\delta^2 S_\mathrm{ren}}{\delta a_{t(0)}^{Ta}(x) \delta \lambda_{(0)}(y)} = 
-\frac{i}{2} (\tilde f^a (\partial)^*+k^a(\partial))\delta^3(x-y)\ 
\end{equation}
and similarly for the others.
In the main text we have opted for the splitting \eqref{split1} because it more directly reflects the loss of boost invariance, and it avoids the non-local expression in \eqref{gaugeinvbis}.

\section{Correlators and Ward identities in a field theory model}

In this Appendix we discuss the field theory model of \cite{Miransky:2001tw,Schafer:2001bq} which shares the same symmetries and pattern of symmetry breaking as the holographic model discussed in this note. In the weak coupling limit, we can compute the two-point correlators at tree level, verify once more the Ward identities but also extract the dispersion relations and the spectrum. 

The model consists in just a complex scalar doublet $\Phi$ of a $U(2)$ global symmetry, with a chemical potential $\mu$ for the abelian $U(1)$ current. The action is
\begin{equation}
S=\int d^3x \left[ (\partial_t +i\mu)\Phi^\dagger (\partial_t-i\mu)\Phi-\partial_u \Phi^\dagger \partial_u \Phi-M^2 \Phi^\dagger\Phi -\lambda(\Phi^\dagger\Phi)^2\right]\ .
\end{equation}
The conserved currents are given by
\begin{equation}
J^a_i=i\Phi^\dagger T^a\partial_i \Phi-i\partial_i \Phi^\dagger T^a \Phi- 2\mu \delta_{ti}\Phi^\dagger T^a \Phi\ .
\end{equation}
One can verify that they satisfy the $U(2)$ current algebra, for any value of $\mu$ (which is real). 

When $\mu^2>M^2$ (i.e. always if $M^2<0$) the theory settles in a vacuum where the global symmetry is broken down to a $U(1)$, for definiteness we choose:
\begin{equation}
\Phi_v= \left(\begin{array}{c} 
0 \\ v
\end{array}\right)\ , \qquad v^2=\frac{\mu^2-M^2}{2\lambda}\ .
\end{equation}
We can now consider the small fluctuations around the vacuum:
\begin{equation}
\Phi=\Phi_v+\left(\begin{array}{c} 
\xi \\ \varphi
\end{array}\right)\ .
\end{equation}
Expanding the action to quadratic order we obtain the following tree-level propagators:
\begin{align}
\langle \xi(\omega,k)\xi^\dagger(-\omega,-k)\rangle &= \frac{i(\omega^2-k^2-2\mu\omega)}{(\omega^2-\omega_{\xi+}^2)(\omega^2-\omega_{\xi-}^2)}\ , \nn
\langle \xi(\omega,k)\xi(-\omega,-k)\rangle &= 
\langle \xi^\dagger(\omega,k)\xi^\dagger(-\omega,-k)\rangle =0\ ,\nn
\langle \varphi(\omega,k)\varphi^\dagger(-\omega,-k)\rangle &= \frac{i(\omega^2-k^2-2\mu\omega-2\lambda v^2)}{(\omega^2-\omega_{\varphi+}^2)(\omega^2-\omega_{\varphi-}^2)}\ , \nn
\langle \varphi(\omega,k)\varphi(-\omega,-k)\rangle &=
\langle \varphi^\dagger(\omega,k)\varphi^\dagger(-\omega,-k)\rangle= \frac{2i\lambda v^2}{(\omega^2-\omega_{\varphi+}^2)(\omega^2-\omega_{\varphi-}^2)}\ ,
\end{align}
with the dispersion relations
\begin{align}
\omega_{\xi\pm}^2&=k^2+2\mu^2\pm2\mu\sqrt{ \mu^2+k^2}\ , \nn
\omega_{\varphi\pm}^2&= k^2+3\mu^2-M^2\pm\sqrt{(3\mu^2-M^2)^2 +4\mu^2k^2}\ ,
\end{align}
which at small momenta $k$ give
\begin{align}
\omega_{\xi+}&=2\mu+\dots\ , \nn 
\omega_{\xi-}&=\frac{k^2}{2\mu}+\dots\ , \nn
\omega_{\varphi+}&=\sqrt{6\mu^2-2M^2}+\dots\ , \nn 
\omega_{\varphi-}&=\sqrt{\frac{\mu^2-M^2}{3\mu^2-M^2}}k +\dots
\end{align}
We see that we have all the variety of massless and massive modes: $\omega_{\varphi-}$ for a type A Goldstone boson with $c_s<1$, $\omega_{\xi-}$ for a type B Goldstone boson, $\omega_{\xi+}$ for its massive partner and 
$\omega_{\varphi+}$ for a Higgs-like massive excitation. Furthermore, we can compute the VEV of the currents and find, e.g., $\langle J_t^3 \rangle=\mu v^2$, which correctly vanishes when $\mu\to 0$.

In the broken symmetry vacuum, the currents can also be expanded at the linear order:
\begin{align}
J^0_i&=\frac{i}{2}v\partial_i(\varphi-\varphi^\dagger)-\delta_{ti}\, \mu v (\varphi+\varphi^\dagger) \ , \nn
J^1_i&=\frac{i}{2}v \partial_i(\xi-\xi^\dagger)-\delta_{ti}\, \mu v (\xi+\xi^\dagger) \ , \nn
J^2_i&=-\frac{1}{2}v \partial_i(\xi+\xi^\dagger)-i\delta_{ti}\, \mu v (\xi-\xi^\dagger) \ , \nn
J^3_i&=-J^0_i\ .
\end{align}
It is then now straightforward to compute the correlators analogous to the ones in \eqref{corrjtjt}--\eqref{corrjuop}. 

Let us for instance compute one such set of correlators:
\begin{align}
\langle J^1_t (\omega,k)\xi(-\omega,-k)\rangle &=\frac{1}{2}v(\omega-2\mu )\frac{i(\omega^2-k^2+2\mu\omega)}{(\omega^2-\omega_{\xi+}^2)(\omega^2-\omega_{\xi-}^2)}\ ,\nn
\langle J^1_u (\omega,k)\xi(-\omega,-k)\rangle &=\frac{1}{2}v k_u\frac{i(\omega^2-k^2+2\mu\omega)}{(\omega^2-\omega_{\xi+}^2)(\omega^2-\omega_{\xi-}^2)}\ ,
\end{align}
which are clearly non-local since they contain the poles for the type B Goldstone boson and its massive partner. However the combination
\begin{equation}
-i\omega \langle J^1_t (\omega,k)\xi(-\omega,-k)\rangle
+
ik_u \langle J^1_u (\omega,k)\xi(-\omega,-k)\rangle  
=
\frac{v}{2}\ 
\end{equation}
is constant, by virtue of a Ward identity similar to the one in \eqref{wilambda}, for the above correlator.

All the other Ward identities can be similarly checked. In this weakly coupled model however, as we have seen, we have access also to the individual correlators and hence to the non-local parts that contain complete information on the spectrum and on the dispersion relations.



  \end{document}